\documentclass[a4paper,fleqn,usenatbib]{mnras}

\usepackage[T1]{fontenc}
\usepackage{ae,aecompl}

\usepackage{graphicx}	
\usepackage{epstopdf}
\usepackage{amsmath}	
\usepackage{amssymb}	
\usepackage{nccmath}

\usepackage{footnote}

\usepackage{soul}

\newcommand{\elias}{KPD19 }

\title[Multiwavelength PSD analysis of NGC 5548]{Multiwavelength power-spectrum analysis of NGC 5548}

\author[Panagiotou et al.]{C. Panagiotou,$^{1}$\thanks{E-mail: Christos.Panagiotou@unige.ch} 
I. E. Papadakis,$^{2,3}$
E. S. Kammoun,$^{4}$
M. Dov\v ciak$^{5}$
\\
\\
$^{1}$Astronomy Department, University of Geneva, Chemin d'Ecogia 16, 1290 Versoix, Switzerland\\
$^{2}$Department of Physics and Institute of Theoretical and Computational Physics, University of Crete, 71003 Heraklion, Greece \\
$^{3}$IA, FORTH, Voutes, GR-7110 Heraklion, Greece \\
$^{4}$Department of Astronomy, University of Michigan, 1085 South University Avenue, Ann Arbor, MI 48109-1107, USA\\
$^{5}$Astronomical Institute of the Academy of Sciences, Bo\v cn\'i II 1401, CZ-14100 Prague, Czech Republic\\
}

\date{Accepted XXX. Received YYY; in original form ZZZ}

\pubyear{2020}

\begin{document}
\label{firstpage}
\pagerange{\pageref{firstpage}--\pageref{lastpage}}
\maketitle

\begin{abstract}

NGC 5548 was recently monitored intensively from NIR to X-rays as part of the STORM campaign. Its disc emission was found to lag behind the observed X-rays, while the measured time lag was increasing with wavelength. These results are consistent with the assumption that short term variability in AGN emission is driven by the X-ray illumination of the accretion disc. In this work, we studied the power spectrum of UV/optical and X-ray emission of NGC 5548, using the data of the STORM campaign as well as previous \textit{Swift} data, in order to investigate the relation between the UV/optical and X-ray variability and to examine its consistency with the above picture. We demonstrate that even the power spectrum results are compatible with a standard disc being illuminated by X-rays, with low accretion rates, but the details are not entirely consistent with the results from the modelling of the ``$\tau$ vs $\lambda$" relation. The differences indicate that the inner disc might be covered by a ``warm corona" which does not allow the detection of UV/optical emission from the inner disc. Finally, we found strong evidence that the UV emission of NGC 5548 is not stationary.

\end{abstract}

\begin{keywords}
galaxies: active -- galaxies: nuclei -- galaxies: Seyfert -- galaxies: individual: NGC 5548
\end{keywords}



\section{Introduction}

It is generally accepted that Active Galactic Nuclei (AGN) are powered by the release of gravitational energy during accretion of matter, in the form of a disc, onto a supermassive black hole (BH). If the disc is geometrically thin and optically thick, it radiates a multi-temperature blackbody emission with a radial temperature profile of $T(r) \propto r^{-3/4}$ \citep[][NT73 hereafter]{1973A&A....24..337S, 1973blho.conf..343N}. AGN are also strong X-ray emitters. Since the  disc temperature is not expected to reach values high enough to account for the observed X-ray emission, X-rays are thought to be produced by the Comptonisation of disc photons by high-energy electrons, located in a hot ($kT_{\text{e}}\sim100$ keV) and optically thin region, which is usually referred to as the ``X-ray corona" \citep[e.g.,][]{1993ApJ...413..507H}. Fast X-ray variability and microlensing studies \citep[e.g.,][]{2016AN....337..356C} suggest that the X-ray source lies within $\sim 10-20 \text{ }R_{\text{g}}$ from the BH, where $R_\text{g}=GM_{\text{BH}}/c^2$  is the gravitational radius ($M_{\text{BH}}$ being the BH mass).

The exact geometry of the disc/corona system is still unclear, and the disc/corona interaction is currently the subject of intense research. Several efforts have been made to investigate the disc/corona link by studying the X-ray spectrum of AGN. In addition, if the X-ray corona is located above the disc and emits isotropically, we expect a strong correlation between the UV/optical and X-ray variability in AGN. After all, if X-rays illuminate the disc, part of the flux will be reflected and the rest will be absorbed, increasing the local temperature of the disc. In the case when the X-rays are variable, the additional thermal disc emission will also be similarly variable, but with a delay, which depends on the disc/X-ray corona geometry. Early efforts \citep[e.g.,][]{1992ApJ...393..113C} were supportive of the thermal reverberation scenario, by detecting a strong correlation between the X-ray and the UV/optical emissions on short time-scales. 

In the last years, a few bright AGN have been observed intensively, and simultaneously, in X-rays and the UV/optical bands, using mainly the {\it Neil Gehrels/Swift} observatory \citep[e.g.,][]{2014MNRAS.444.1469M, 2015ApJ...806..129E, 2018MNRAS.480.2881M, 2019ApJ...870..123E}. The first target of such an intense monitoring was NGC 5548, a typical Seyfert galaxy located at z=0.01717 \citep{1991rc3..book.....D}. It hosts a BH with a mass of $M_{\text{BH}}=5\cdot 10^7 M_\odot$ \citep{2015PASP..127...67B}, which accretes at a moderate rate of around 5\% of the Eddington limit \citep{2016ApJ...821...56F}. \cite{2015ApJ...806..129E} and  \cite{2016ApJ...821...56F} used the NGC 5548 {\it Swift}, {\it HST} and ground based data, that were collected within the framework of the ``Space Telescope and Optical Reverberation Mapping" \citep[STORM, ][]{2015ApJ...806..128D} campaign, and they found that UV/optical light curves are very well correlated, but with a delay (or ``time lag", $\tau$)  which is increasing with the wavelength as $\tau \propto \lambda^{4/3}$. This is in agreement with the predictions in the case of a standard accretion disc being illuminated by a compact, point--like corona (i.e., in the case of the so-called ``lamp-post" geometry). They also found that the normalisation of the $\tau-\lambda$ relation is larger than expected (for the assumed accretion rate of NGC 5548) and that the correlation between the X-rays and the UV/optical bands was rather moderate. \citet{2017ApJ...835...65S} used the same data and inferred the ``driving" light curve for the observed UV/optical variations, which was not in agreement with the observed X-ray light curve. On the other hand, \citet*[][KPD19 hereafter]{2019ApJ...879L..24K} showed that a point-like X-ray source and a ``standard" NT73 disc emitting at around 1 per cent of the Eddington limit can explain the amplitude of the time-lags vs wavelength relation in NGC 5548; as long as the X-ray source is located at a height larger than about 40-60 $R_\text{g}$ above the BH. 

So far, the NGC 5548 data have been studied within the context of cross-correlation analysis, only. However, the superior quality of this data set (in terms of sampling rate, duration, and signal-to-noise ratio) allows for a proper power-spectral density (PSD or simply power-spectrum) analysis, from X-rays up to near infrared. Power spectrum analysis provides information about the amplitude of the variability components on various time-scales, and the way the PSD changes with wavelength can put strong constraints on any model for the UV/optical variability of AGN. 

In this work, we present the results from the PSD analysis of the multi-wavelength light curves that were obtained within the STORM campaign of NGC 5548. To the best of our knowledge, this is the first time that such a PSD study is performed, over (almost) all the energy bands where we can detect direct emission from an AGN, using contemporaneous light curves.  Section \ref{sec:data_sample} describes the used data set, while the performed PSD analysis is presented in Sect. \ref{sect:psd} and the evolution of PSD amplitude with wavelength is studied in Sect. \ref{sec:frs}. We discuss our main results and conclusions in Sect. \ref{sec:discus}

\section{Data Sample}
\label{sec:data_sample}

NGC 5548 has been monitored extensively by {\it Swift} since its launch. We considered two periods of observations during which the source was monitored with high cadence: from 6384 to 6548 (in units of HJD-2,450,000), and from 6613 to 6876 (periods P1 and P2, hereafter).  We did not consider observations before P1 because of their lower cadence. During a shorter period within the P2 time interval, NGC 5548 was the target of the STORM campaign, which lasted for almost 200 days. NGC 5548 was simultaneously observed by {\it HST} and {\it Swift}, as well as by several ground-based telescopes, from near infrared to X-rays. The source was intensively monitored, with a cadence of up to two observations per day in some wavebands. 

Using the {\it Swift/XRT} data and the online products building tool\footnote{www.swift.ac.uk/user\_objects/} \citep{2009MNRAS.397.1177E}, we constructed the hard X-ray (HX) ligtcurve from 2 to 7 keV, during both the P1 and P2 periods. We did not consider a lower energy limit so that we are not affected by any soft band absorption variations \cite[e.g.,][]{2014Sci...345...64K}. The data were acquired in the photon counting mode. The mean count rate was 0.26 in P1 and 0.35 during P2.

Regarding the UV and optical light curves,  we chose data in bands with long and densely sampled light curves, with the smallest possible error, and with filters that have relatively narrow and non-overlapping effective areas.  Table \ref{tab:filtra_infos} lists the UV/optical bands that we considered and some observational details for each band. 

We used the far UV, $ \lambda 1158$ and $\lambda 1479$ {\it HST} continuum light curves from  \cite{2016ApJ...821...56F}. The {\it HST} observations mark the core of the STORM campaign. We used the UVW1 and UVW2 light curves, during periods P1 and P2, from \cite{2015ApJ...806..129E}, who have reduced all the {\it Swift/UVOT} observations of NGC 5548, taken until 2014 August 6. The rest of the UV/optical light curves were taken from \cite{2016ApJ...821...56F}, who provide a detailed description of the data reduction method. 

\begin{table*}
	\centering
	\caption{Energy bands and filters used in this work. $N_{obs}$ and $\Delta t$  denote the number of observations and the average time between two successive observations, respectively (numbers in parantheses refer to the interpolated light curves). The start and end time (in units of HJD-2,450,000) as well as the total duration of each light curve are given in the next two columns. The last column lists the flux of the host galaxy in units of $10^{-15}\text{ergs/s/}\text{cm}^2/ \text{\AA}$, as taken by \protect \cite{2016ApJ...821...56F}.}
	\label{tab:filtra_infos}
	\begin{tabular}{lcccccc} 
		\hline
		{Filter}  &  $\lambda$ (\AA)    &$N_{obs}$    &$\Delta t$ (days)    &Start/End Date           & T (days)   &$f_{\lambda,host}$  \\
		\hline
		HX/P1     &                     & 146 (163)   &  1.13 (1.0)            & 6384.09/6547.59       & 163.5          &    --       \\
	  HX/P2     &                     & 330 (262)   &  0.80 (1.0)            & 6613.76/6875.61       & 261.9          &    --       \\
		H1        & 1158                & 171 (350)   &  1.03 (0.5)            & 6690.61/6865.92       & 175.3          &    --       \\
		H3        & 1479                & 171 (350)   &  1.03 (0.5)            & 6690.65/6865.94       & 175.3          &    --        \\
		W2/P1     & 1928                & 228 (327)   &  0.72 (0.5)            & 6384.01/6547.62       & 163.6          &    --        \\
		W2/P2     &                     & 321 (524)   &  0.82 (0.5)            & 6613.75/6875.61       & 261.9          &    --        \\
		W1/P1     & 2600                & 134 (328)   &  1.23 (0.5)            & 6383.99/6547.61       & 163.6          &    --        \\  
		W1/P2     &                     & 279 (328)   &  0.59 (0.5)            & 6666.07/6830.42       & 164.3          &    --        \\
		SDSS u    & 3467                & 164 (212)   &  1.30 (1.0)            & 6684.78/6896.38       & 211.6          & $1.16 \pm 0.02$    \\
		Johnson B & 4369                & 180 (224)   &  1.25 (1.0)            & 6645.64/6869.31       & 223.7          & $2.88 \pm 0.05$           \\
		Johnson V & 5404                & 497 (576)   &  0.58 (0.5)            & 6645.62/6933.62       & 288.0          & $4.79 \pm 0.10$           \\
		SDSS r    & 6176                & 203 (211)   &  1.04 (1.0)            & 6684.78/6895.36       & 210.6          & $5.76 \pm 0.12$           \\
		SDSS i    & 7648                & 208 (211)   &  1.02 (1.0)            & 6684.78/6895.36       & 210.6          & $5.33 \pm 0.10$           \\
		SDSS z    & 9157                & 212 (211)   &  1.00 (1.0)            & 6684.78/6895.36       & 210.6          & $5.00 \pm 0.08$           \\
		\hline  
	\end{tabular}
\end{table*} 

\section{Power Spectrum Analysis}
\label{sect:psd}
\subsection{The PSD estimation}
Assuming a discrete time series $f(t_i)$ with N equidistant points, the typical way to estimate  its PSD is by calculating the periodogram:

\begin{equation}
  \centering
      P(\lambda,\nu_j) = \frac{2\Delta t}{N} \left\lvert \large\sum_{k=1}^{N} (f(\lambda,t_k)-\bar{f}_{\lambda})e^{2\pi i \nu_j t_k}  \right\rvert ^2,
\label{eq_psd}
\end{equation}

\noindent where $f(\lambda,t_k)$ is the source count rate (or flux) at wavelength $\lambda$ and time $t_k$, $\bar{f}_{\lambda}$ is the light curve mean, and $\Delta t$ is the bin size. The PSD is calculated at the discrete frequency values $\nu_j = \dfrac{j}{t_N - t_1}$, where $j=1,2,...,N/2$. It is customary to divide $f(t_i)$ by the light curve mean, in which case the periodogram is an estimate of the variability amplitude normalised to the mean flux. We therefore divided the periodograms by $(\bar{f}_{\lambda}-f_{\lambda,host}){^2}$, where $f_{\lambda,host}$ is the host galaxy flux in each filter, listed in the last column of Table \ref{tab:filtra_infos} (in this way, the variability amplitude is properly normalized to the intrinsic AGN emission flux). Since the periodogram calculation requires evenly sampled data, we produced new equidistant light curves by interpolating the observed light curves as follows. 

We chose an initial time, $t_0$, slightly larger than the time of the first observation, and a time step, $\Delta t_{\rm inter}$, which was close to the average cadence of the observed light curve. Then, at each time $t_k = t_0 + k \cdot \Delta t_{\rm inter}$ (with $k=0,1,2,...$) we estimated the flux using linear interpolation between the nearest previous and succeeding flux points, and we added random noise equal to the average error of all the points in the original light curve. The number of points and the bin size of the interpolated light curves are also listed in Table \ref{tab:filtra_infos} (values in parentheses in the third and fourth column, respectively). 

\begin{figure}
\centering
\includegraphics*[height=240pt, width=\columnwidth]{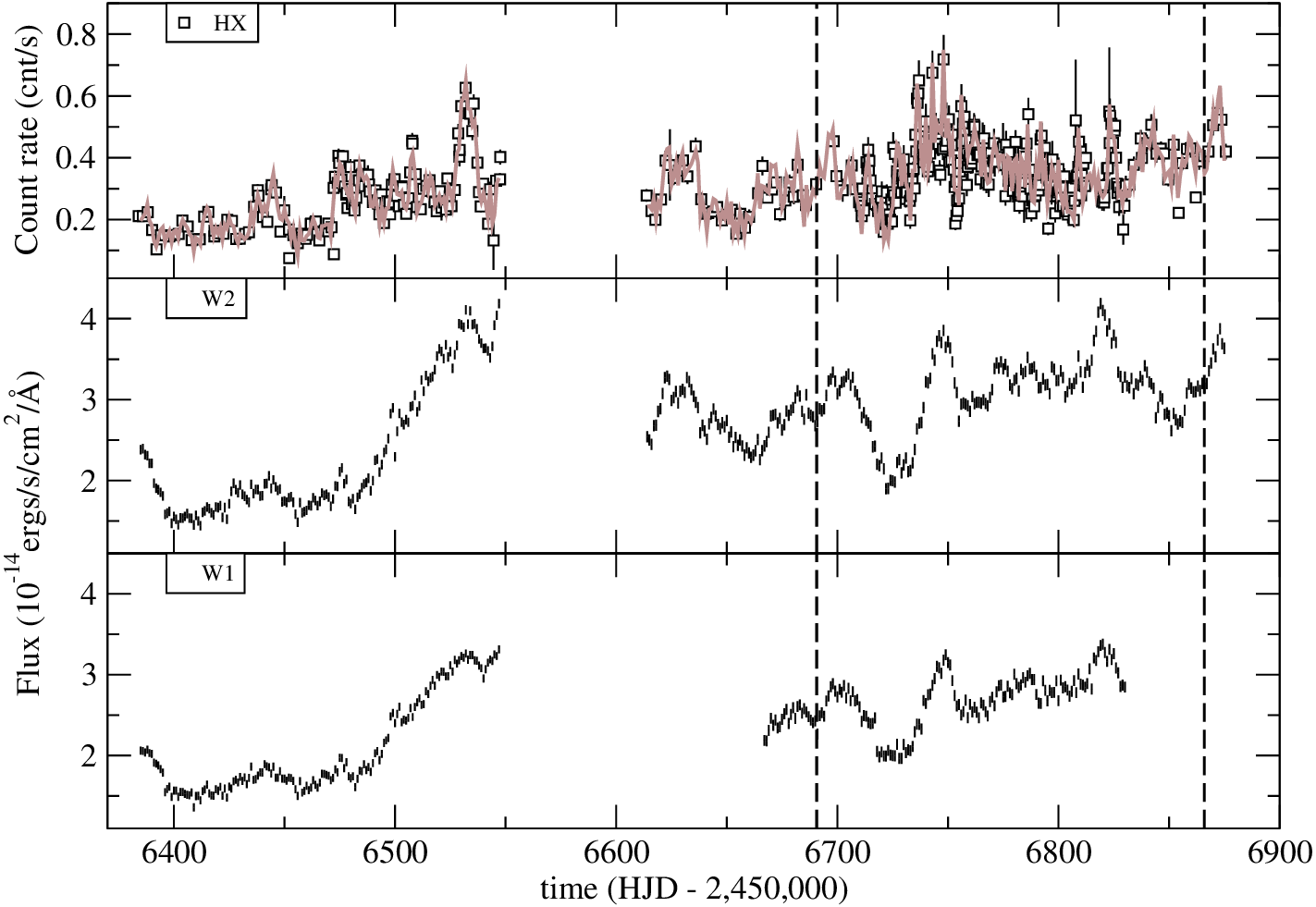}
 \caption[10]{Interpolated X-ray, W2 and W1 light curves. Open squares in the top panel indicate the observed data. The vertical dashed lines indicate the start/stop dates of the HST observations, plotted in Fig. \ref{fig:lc_4}.}
 \label{fig:lc_1}
\end{figure}

\begin{figure}
\centering
\includegraphics*[height=380pt, width=\columnwidth]{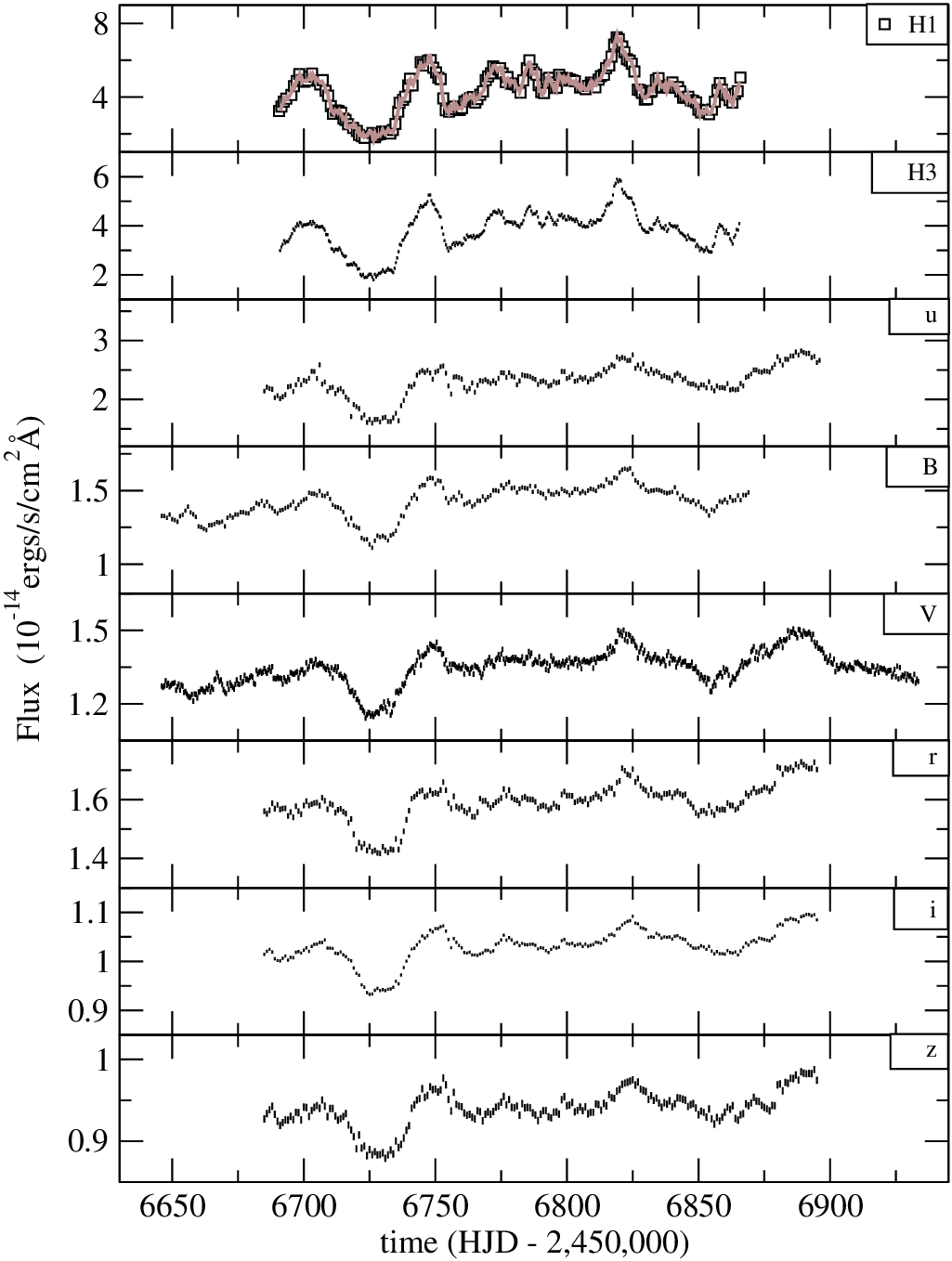}
 \caption[10]{The interpolated {\it HST} (H1 and H3 bands), and UV/optical light curves. Open squares in the top panel indicate the observed light curve.}
 \label{fig:lc_4}
\end{figure}

The interpolated light curves are plotted in Fig. \ref{fig:lc_1} and \ref{fig:lc_4}. They are almost identical to the observed light curves and reproduce well all the observed variations. To demonstrate this, in the top panel of the aforementioned figures we plot both the observed and the interpolated light curve (open squares and solid lines, respectively). The vertical, dashed lines in Fig.\,\ref{fig:lc_1} indicate the period when the {\it HST} observations were taken (the HST light curves are plotted in Fig.\,\ref{fig:lc_4}, and they indicate the time span of the STORM campaign). 


\begin{figure}
  \centering
  \includegraphics*[height=240pt, width=\columnwidth]{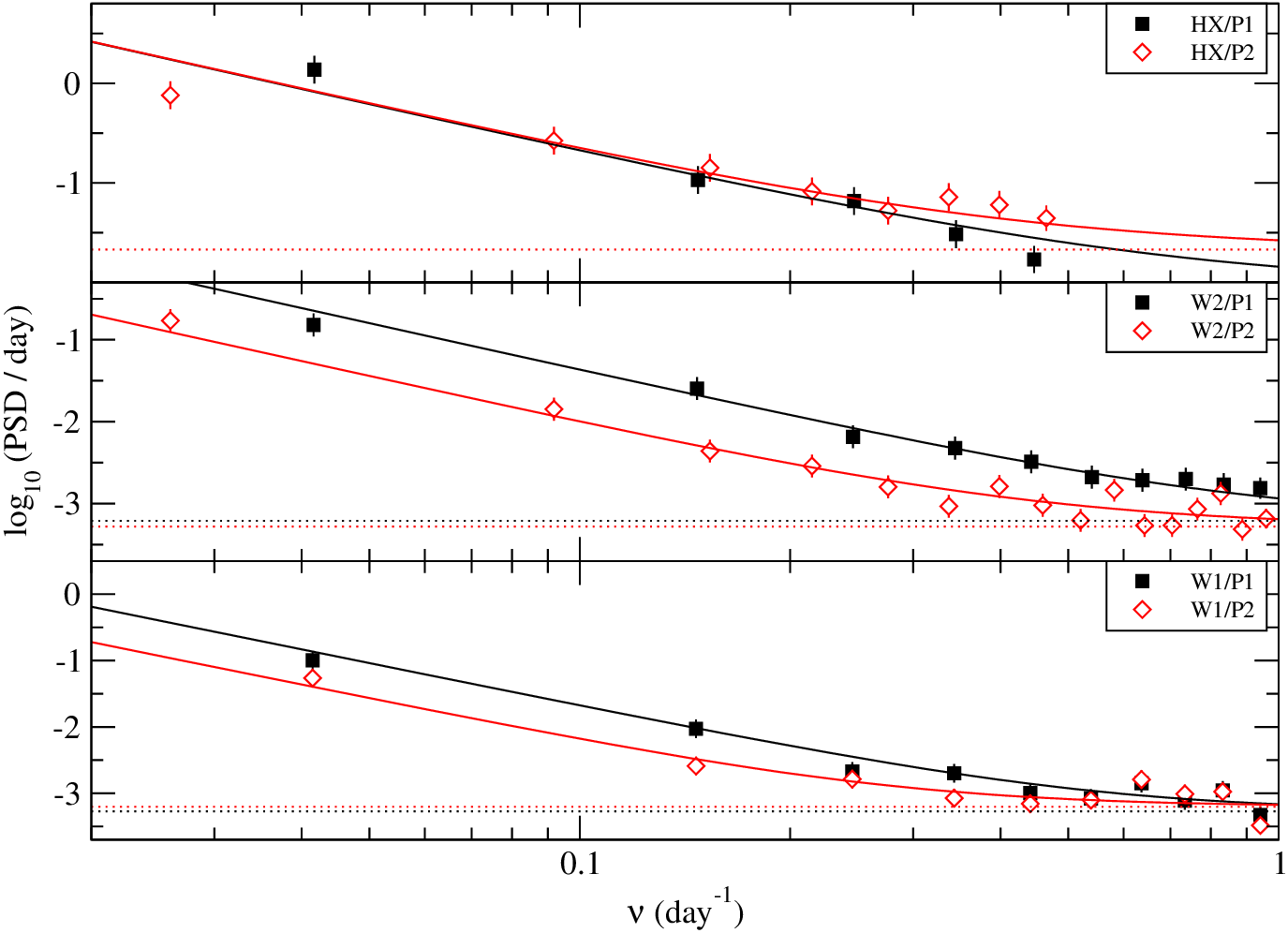}
  \caption[10]{The hard X-ray, W2 and W1 power spectra (from top to bottom), estimated with the light curves from the P1 and P2 periods (black and red markers, respectively, in each panel). The solid lines correspond to the best-fitting models and the dotted lines indicate the constant Poisson noise level. }
  \label{fig:psd_1}
\end{figure}

\begin{figure}
  \centering
  \includegraphics*[height=240pt, width=\columnwidth]{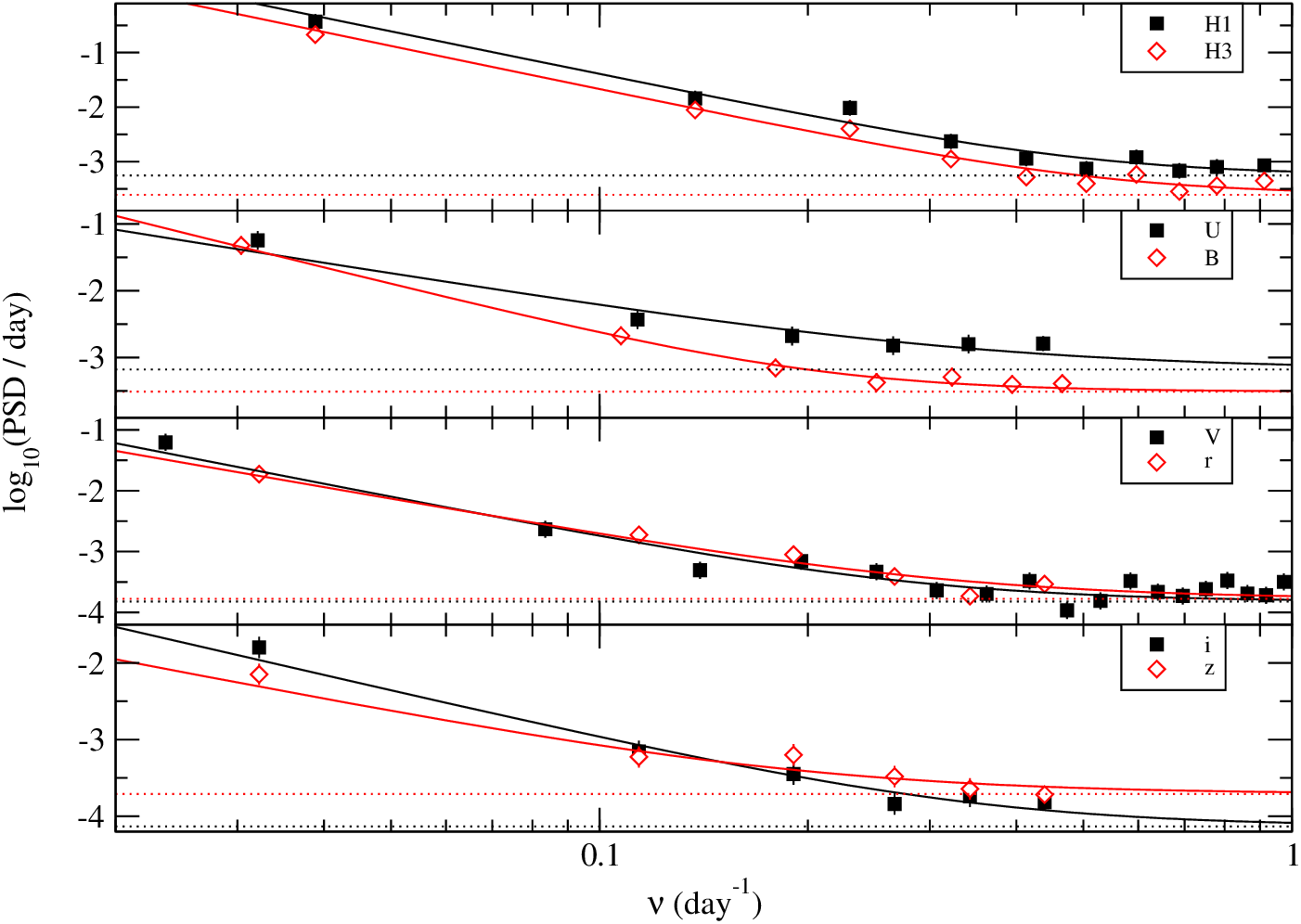}
  \caption[10]{Same as Fig.\,\ref{fig:psd_1} for the rest UV/optical bands.}
  \label{fig:psd_2}
\end{figure}

Finally, we binned the periodogram of the interpolated light curves (in the log--log space), using a bin size of 16, following the method proposed by \cite{1993MNRAS.261..612P}. The estimated PSDs are plotted in Fig. \ref{fig:psd_1} and \ref{fig:psd_2}. As it is shown in Fig. \ref{fig:psd_1}, two different periodograms were calculated for each of HX, W2 and W1 light curves, one for the P1 and one for the P2 light curve in each band, respectively.  


\subsection{PSD model fit results}
\label{sec:psd_fit}

We fitted all the PSDs with a power-law model of the form

\begin{eqnarray}
\label{eq:psd_model}
  \centering
      P_m(\lambda, \nu) = A_{\lambda} \left ( \frac{\nu}{0.1} \right )^{-\alpha(\lambda)} + C_{\lambda},
\end{eqnarray}

 \noindent where frequencies are measured in  units of $\text{day}^{-1}$, the PSD normalisation, $A_{\lambda}$, is defined as the PSD value at $\nu=0.1 \text{ }{\rm day}^{-1}$, and $C_{\lambda}$ is the constant power due to Poisson noise\footnote{We kept $C$ frozen to its predicted value, that is $2\Delta t\bar{\sigma}^2/(\bar{f}_{\lambda}-f_{\lambda,host}){^2}$, where $\bar{\sigma}$ is the average error of the points in the original light curves}. We considered model fits to be statistically acceptable if the $p-$value (using the $\chi^2$ statistic) is larger than 1\%. We compared different model fits using the $F-$test; models with larger number of free parameters give a significantly better fit to the PSD if the corresponding null hypothesis probability is less than 1\%.

First we fitted the HX P1 and P2 PSDs together, keeping the normalisation and the slope the same for both spectra. The resulting fit was good, with $\chi^2=14.2$ for 11 degrees of freedom (dof; $P_{\text{null}} = 22.2\%$).The quality of the fit did not improve significantly when we let the normalisation and/or the slope to vary between the two periods. The best-fitting models are plotted in the top panel of Fig. \ref{fig:psd_1}.

Then we fitted the W2 and W1 PSDs. Similarly to the X-rays, we fitted the power spectra of the P1 and P2 light curves together. The fit was poor when both the normalisation and the slope were linked between the two periods ($\chi^2/\text{dof} = 81.3/24$ and $38.7/18$ for W2 and W1, respectively). The fit was significantly improved when the normalisation was let to be different while the slope was still kept linked ($\chi^2/\text{dof} = 20.8/23$ and $25.2/17$). Allowing for different slopes between the two periods does not provide a significantly better fit neither to the W2 nor to the W1 power spectrum. The best-fitting models are plotted in Fig. \ref{fig:psd_1}. The difference in the PSD normalisation is apparent just by looking at the PSDs, and it is more prominent in the W2 power spectra. Looking at the W2 and W1 light curves (Fig. \ref{fig:lc_1}) one may notice that the min-to-max flux variations during the P1 period are larger than the respective variations in the P2 light curves. This difference in the variability amplitude is the reason for the difference in the best-fitting PSD normalisation between the two periods. The variability difference was further supported by a comparison of the fractional variability amplitudes, $F_\text{var}$ \citep[e.g.,][]{2003MNRAS.345.1271V}. It was found that $F_\text{var, W2}=0.335 \pm 0.002$, $F_\text{var, W1}=0.231 \pm 0.002$ during P1 and $F_\text{var, W2}=0.165 \pm 0.001$, $F_\text{var, W1}=0.133 \pm 0.002$ during P2. This is in agreement with our PSD analysis results, which show that the difference in variability amplitude is due to PSD amplitude variations, with a constant slope.

\begin{figure}
  \centering
  \includegraphics*[height=200pt, width=\columnwidth]{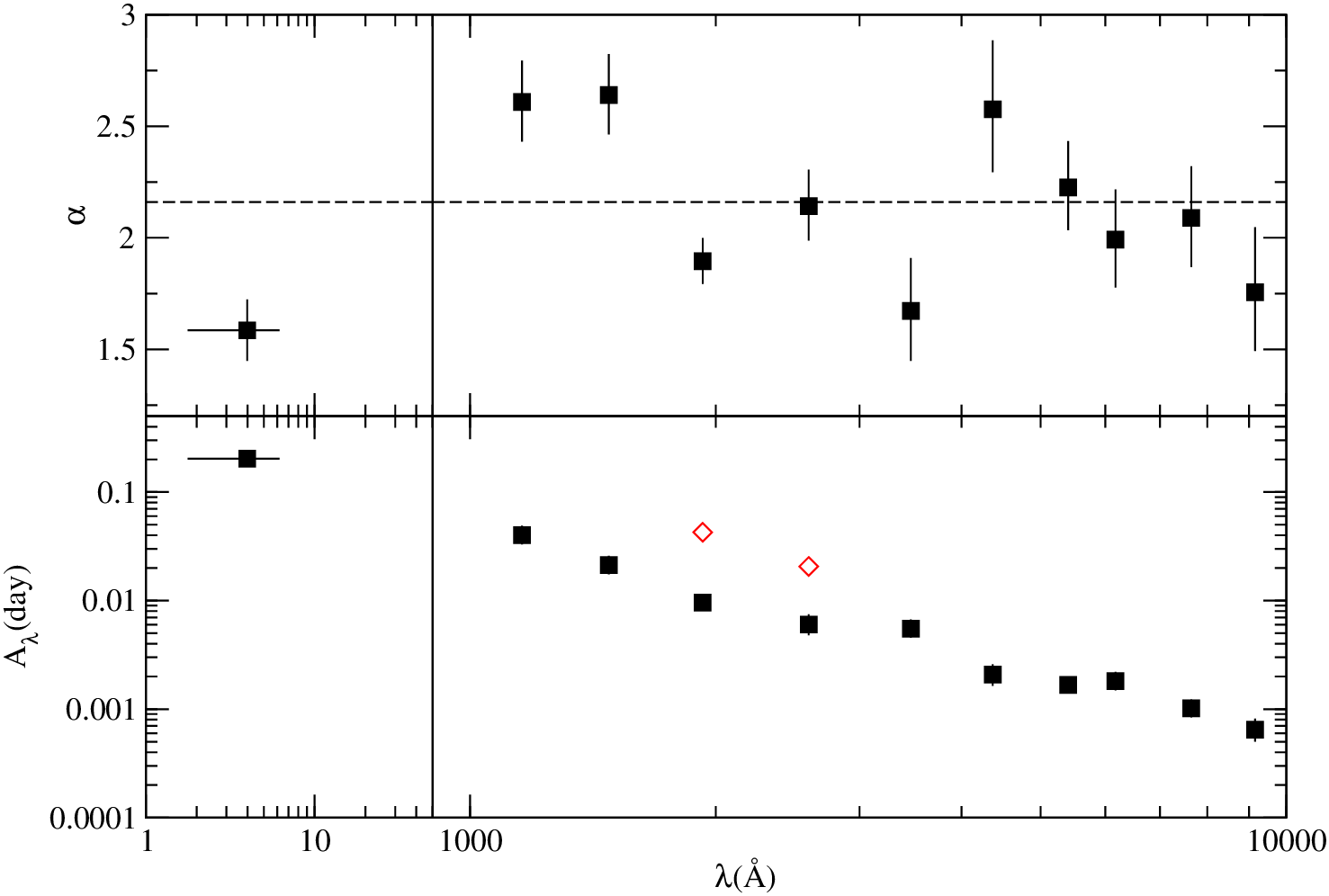}
  \caption[10]{Best-fitting slope (top) and normalisation (bottom panel) of all UV/optical and HX (left) PSDs. The x error of HX points denotes the energy range of the used light curve. The dashed line in the top panel denotes the average value and the red open diamonds in the lower panel correspond to the best-fitting values for W2 and W1 during the P1 observation period. The errors are not distinguishable because of their small values.}
  \label{fig:best_fit}
\end{figure}

The PSDs in the other bands are well fitted by the power-law model. The best-fitting results are listed in Table \ref{tab:best-fit}, and the corresponding models are plotted in Fig. \ref{fig:psd_2}. The best-fitting slope and normalisation of all the PSDs are plotted as a function of wavelength in Fig. \ref{fig:best_fit}. The dashed line in the top panel indicates the mean slope, $\bar{\alpha} = 2.16 \pm 0.11$ of the UV/optical PSDs. The UV/optical data are not entirely consistent with the hypothesis of a common slope at all bands ($\chi^2/dof = 29.3/9$,  $P_{null} = 0.06\%$). This is probably due to the slope of the HST PSDs being somewhat steeper than the PSD slope in the other bands. Indeed, the mean PSD slope without the HST data is  $2.04 \pm 0.10$. It is fully consistent with the previous value, but now the hypothesis of the same PSD slope  is consistent with the data ($\chi^2/dof = 10.31/7$,  $P_{null} = 17\%$). 

The best-fitting results plotted in Fig. \ref{fig:best_fit} summarise well the results from the PSD analysis: the main difference between the PSDs in the UV/optical bands is their normalisation. The PSD slope is more or less consistent with a value equal to about $2-2.5$ in all wavebands, which is significantly higher than the slope of the HX PSD. However, the PSD normalisation decreases with increasing wavelength (bottom panel in Fig \ref{fig:best_fit}). This is the main difference between the power spectra of the light curves in the various optical/UV bands. For that reason, we will study below the dependence of the PSD normalisation on wavelength, and we will compare it with model predictions within the context of disc X-ray illumination models. 

\begin{table}
	\centering
	\caption{The best fit power-law slope, $\alpha$, and normalisation, $A_{\lambda}$, of the PSD in the various wavebands we considered in this work. All the errors correspond to 1-$\sigma$ error.}
	\label{tab:best-fit}
	\begin{tabular}{rccccccc} 
		\hline
		{Filter}   &$\alpha$              & A$_{\lambda}$                 & $\chi^2$ /dof        &$P_{null}$     \\
		          &                               & ($10^{-3} \text{ }\mathrm{day}$)     &                     &  (\%)             \\
		\hline
		HX        &  ${1.59} \pm 0.14$            &  ${203.3}^{+28.11}_{-24.84}$   &   14.2/11    &   22.2        \\
		H1        &  ${2.61}^{+0.19}_{-0.18}$     &  ${40.21}^{+8.76}_{-7.21}$     &   9.1/8      &   33.2        \\ 
    H3        &  ${2.64} \pm 0.18$            &  ${21.26}^{+4.61}_{-3.80}$     &   8.4/8      &   39.8        \\ 
    W2/P1     &  ${1.90} \pm 0.10$            &  ${42.64}^{+7.85}_{-6.63}$     &   20.79/23   &   59.4        \\ 
      /P2     &                               &  ${9.57}^{+1.53}_{-1.33}$      &              &               \\ 
    W1/P1     &  ${2.14}^{+0.16}_{-0.15}$     &  ${20.62}^{+3.56}_{-2.94}$     &   25.21/17   &    9.0        \\ 
      /P2     &                               &  ${6.02}^{+1.47}_{-1.23}$      &              &               \\ 
    u         &  ${1.67}^{+0.24}_{-0.22}$     &  ${5.52}^{+1.17}_{-0.98}$      &   4.4/4      &   35.9        \\ 
    B         &  ${2.58}^{+0.31}_{-0.28}$     &  ${2.08}^{+0.51}_{-0.44}$      &   1.1/5      &   95.2        \\ 
    V         &  ${2.23}^{+0.21}_{-0.19}$     &  ${1.67}^{+0.29}_{-0.26}$      &   28.0/16    &    3.2        \\ 
    r         &  ${1.99}^{+0.22}_{-0.21}$     &  ${1.81}^{+0.38}_{-0.32}$      &   4.6/4      &   33.2        \\ 
    i         &  ${2.09}^{+0.23}_{-0.22}$     &  ${1.02}^{+0.21}_{-0.18}$      &   2.9/4      &   57.0        \\ 
    z         &  ${1.76}^{+0.29}_{-0.26}$     &  ${0.65}^{+0.17}_{-0.14}$      &   3.4/4      &   49.2        \\ 

		\hline  
	\end{tabular}
\end{table}

\section{The ``Fourier-Resolved" Spectrum}
\label{sec:frs}

Any stationary random process can be represented as the sum of sinusoidal functions with a frequency between zero and infinity \citep[e.g.][]{nla.cat-vn2888327}. In fact, the norm squared of these components at frequency $\omega=2\pi\nu$ is equal to PSD$_{\rm intr}(\omega) d\omega$, where PSD$_{\rm intr}(\omega)$ is the intrinsic power spectrum of the process.  If the power-spectrum is known in many wavebands, the distribution of $\sqrt{{\rm PSD}_{\lambda, {\rm intr}}(\omega) d\omega}$ as a function of wavelength, $\lambda$, is the ``Fourier-resolved" spectrum (FRS) of the source at frequency $\omega$. FRS analysis has already been used to study the X-ray variability in X-ray binaries \citep[e.g.,][]{1999A&A...347L..23R, 2003A&A...410..217G} and AGN \citep[e.g.,][]{2007ApJ...661...38P}, but not in the study of the UV and optical variability of AGN.

The shape of the FRS spectra will depend {\it only} on the variable components in the UV/optical bands. Therefore, the Fourier-resolved spectra can reveal the physical component that is responsible for the observed variations in the light curves \citep[see, for example, the discussion in Appendix of][]{2007ApJ...661...38P}. Furthermore, by  studying the FRS in various frequencies, we can investigate if there are different physical components that operate at different time-scales; as these components may show up only in low or high frequency FRS, depending on whether the corresponding mechanism operates on long or short time-scales, respectively. 

Since the slope of the UV and optical PSDs does not change with the wavelength, we decided to estimate the FRS at one temporal frequency only, because its shape will be the same at any other frequency over the sampled frequency range. We chose the frequency $\nu = 0.1$ day$^{-1}$ because it is well within the probed frequency range and the power spectrum in this frequency is well above the Poisson noise level (in all wavebands). In addition, the PSD value at $\nu = 0.1 \text{ }{\rm day}^{-1}$ is equal to the power-law model normalisation, as defined by eq.\,(\ref{eq:psd_model}). We therefore used the best-fitting  PSD amplitudes listed in Table \ref{tab:best-fit}, and we computed the UV/optical FRS of NGC 5548 as follows:

\begin{eqnarray}
\label{eq:frs_def}
  \centering
      \mathcal{R}_{\lambda}(\nu) = (\bar{f}_{\lambda}-f_{\lambda,host}) \sqrt{{A_{\lambda}} \cdot \nu}, \hspace{0.3 cm}({\rm ergs\hspace{0.1 cm}s}^{-1} {\rm cm}^{-2} \text{\AA}^{-1}).
\end{eqnarray}

\begin{figure}
  \centering
  \includegraphics*[height=240pt, width=\columnwidth, clip]{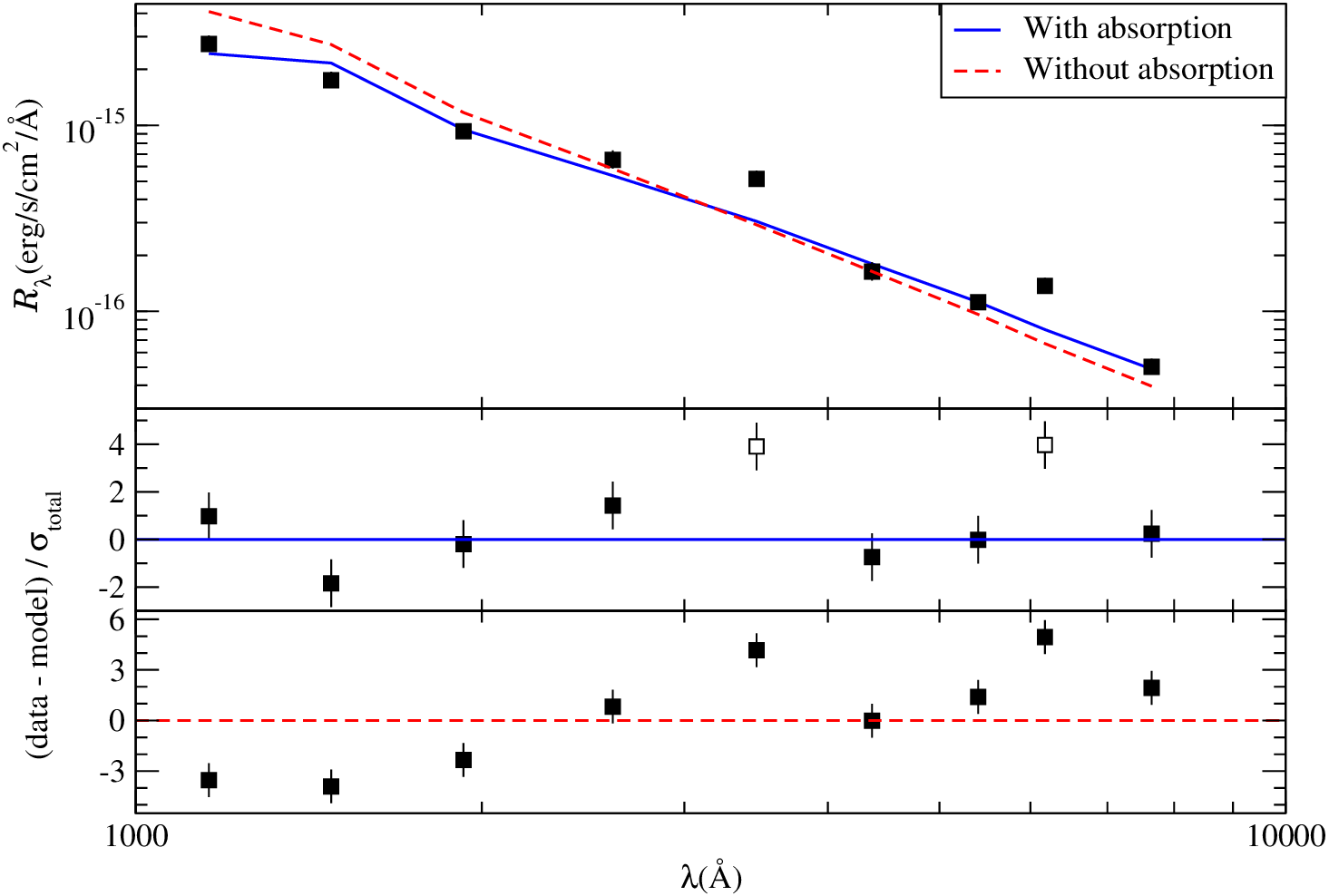}
  \caption[10]{The UV/optical FRS of NGC 5548 at $\nu=0.1$ day$^{-1}$. The red dashed line in the top panel corresponds to the initial best-fitting model assuming X-ray illumination of the disc, while the corresponding residuals are plotted in the bottom panel. The blue solid line of the top panel indicates the final best-fitting model, where absorption of the host galaxy was taken into account and the {\it u}, {\it r} bands were excluded; it corresponds to best-fitting $h=5 \text{ }R_g$ and $\dot{m}_{\rm Edd}=0.5$\%. The middle panel plots the residuals of this fit. The {\it u} and {\it r} filters are plotted for consistency as open squares.}
  \label{fig:frs}
\end{figure}

\noindent The multiplication by  $(\bar{f}_{\lambda}-f_{\lambda,host})$ is necessary to estimate the variability amplitude in physical units, since the PSD was normalised, originally, to the average intrinsic flux of each band. In the case of the W1 and W2 PSDs, we used the P2 best-fitting normalisation values because the P2 light curves coincide with the light curves in the other bands.  The resulting FRS is plotted in the top panel of Fig. \ref{fig:frs}. It decreases at longer wavelengths. Since the PSD slope does not depend on wavelength, this result implies that the total flux of the variable component in each band should decrease with increasing wavelength.

\subsection{Modelling the FRS}
\label{sec:frs_model}

The FRS plotted in Fig. \ref{fig:frs} follows a power-law shape, but a power-law model does not fit the data well ($\chi^2=38.9$ for 7 dof). The power-law model provides an acceptable fit to the data if we ignore the FRS measurements in the $u$ and $r$ bands ($\chi^2=4.6$ for  5 dof;  see the discussion below for the reasons we may wish to exclude these bands). The best fit slope is $-2.12\pm 0.08$. 

One possible explanation as to why the FRS has a power-law shape is if the observed UV/optical variations are due to temperature variations in the inner disc. Figure\, \ref{fig:frs_bb} shows the model FRS in the case of a black body (BB) which varies in temperature (around  a mean of $10^5$ K). The FRS at wavelengths longer than 1000 \AA\, has a power-law shape, but it is steeper than the observed FRS. In fact, we fitted eq. \ref{eq:frs_bb} to the data shown in Fig.\,\ref{fig:frs}, and we could not get a good fit. 

In the case of an X-ray illuminated disc,  the variable UV/optical emission at any given time can be visualised as the sum of thermal emission from various parts of the disc, each with a different temperature, which varies with time. The combination of many variable BB components, that do not vary simultaneously in the same way, may produce a power-law FRS with slope of $-2$. We investigate this possibility below in more detail.

\subsubsection{Testing the  disc X-ray irradiation hypothesis}

In the case of X-ray illumination of the disc, the total disc emission at wavelength $\lambda$ is given by,

\begin{eqnarray}
\label{eq:convolution}
  \centering
      f_{disc}(\lambda, t) =  f_{NT}(\lambda)+\int_{0}^{t} \Psi_\lambda(t-t') \cdot f_X(t') dt',
\end{eqnarray} 

\noindent where $f_{NT}(\lambda)$ is the NT73 disc flux (which we assume does not vary in time). The integral in the right-hand side of eq. (\ref{eq:convolution}) gives the disc flux due to X-ray illumination. $f_{X}(t')$ is the variable X-ray flux, and the function $\Psi_{\lambda}(t)$ is the so-called ``disc response" at wavelength $\lambda$, which depends on the geometry of the disc/corona configuration. The  variable disc emission is the weighted linear combination of past values of the X-ray emission, with the weights being determined by $\Psi_{\lambda}(t)$. The linearity of the system holds as long as the response does not depend on the input, that is if the shape and amplitude of the responce function does not change with time.

If eq.\,(\ref{eq:convolution}) holds, then standard time series analysis theory predicts a simple relation between the output and the input power-spectra \citep[e.g. \S 4.12 in ][]{nla.cat-vn2888327}, 

\begin{eqnarray}
\label{eq:conv_fourier}
  \centering
      P_{\lambda}(\nu) =  | \Gamma_{\lambda}(\nu) |^2 \cdot P_X(\nu),
\end{eqnarray}

\noindent  where $P_{\lambda}(\nu)$ is the power spectrum of the light curve at wavelength $\lambda$, $P_X(\nu)$ is the X-ray power spectrum, and $\Gamma_\lambda$ is the Fourier transform of the response function, known as the "transfer function" of the system,

\begin{eqnarray}
\label{eq:transfer_function}
  \centering
      \Gamma_{\lambda}(\nu) =\int_{-\infty}^{+\infty} \Psi_{\lambda}(t)e^{-i2\pi\nu t}dt.
\end{eqnarray}

\noindent Equation (\ref{eq:conv_fourier}) holds at all frequencies. If we know the ``input" PSD, and the transfer function of the system, we can predict the power spectrum of the ``output" signal. For example, if the input power spectrum is the X-ray PSD, then,

\begin{eqnarray}
\label{eq:frs_transfer}
  \centering
      \mathcal{R}_{\lambda,{\rm mod}}(\nu) = |\Gamma_{\lambda}(\nu) |\sqrt{P_X(\nu) \cdot \nu}.
\end{eqnarray}

\noindent This equation can be used to test any transfer function, that is any disc/corona geometry and physical conditions of the system, as long as the power spectrum of the input source is assumed (or known), and the model transfer function is determined. 

We considered the response functions of \elias to compute the model FRS. \elias modelled the disc response in the lamp-post geometry, for various accretion rates and X-ray corona heights, for a non-rotating and a maximally rotating BH of mass $5\times 10^7$ M$_{\odot}$. They assumed a standard NT73 accretion disc and they computed the disc response taking into account all relativistic effects in the light propagation from the X-ray source to the disc and to the observer. They also  considered the ionization profile of the disc when computing the X-ray reflection spectrum, which is important, as it sets the amount of the energy absorbed by the disc for a given X-ray flux. 

We computed the respective transfer functions using  eq.\,(\ref{eq:transfer_function}), and we used the best-fitting, power-law model parameters for the X-ray PSD (listed in Table\, \ref{tab:best-fit}) to compute the model FRS with eq.\,(\ref{eq:frs_transfer}). We computed 64 model FRS (for the eight accretion rate and eight corona heights that \elias considered), for spin parameter 0 and 1. Since \elias  considered different filters than the filters analysed in this work, we computed $\mathcal{R_{\lambda,\rm mod}}$ by linear interpolation, when necessary. We also excluded the $z$ filter from the following analysis, because its central wavelength is outside the range examined by \elias. Moreover, we considered the effects of Galactic interstellar reddening to the model FRS \cite[as suggested by][]{2007ApJ...661...38P}, assuming the reddening curves of \cite{1989ApJ...345..245C}, $E(B-V)=0.017$ mag \citep{2011ApJ...737..103S}, and $R_V=A_V/E(B-V)=3.1$. 

We computed the $\chi^2$ between the predicted and the observed FRS as follows,

\begin{eqnarray}
\label{eq:chi_square}
  \centering
      \chi^2 = \sum_\lambda \left [ \frac{\mathcal{R}_\lambda - \mathcal{R}_{\lambda,\rm mod}}{\sigma_{\lambda, total}}  \right ]^2,
\end{eqnarray}

\noindent where $\sigma_{\lambda, total} = \left ( \sigma_{\mathcal{R}_\lambda} ^2 + \sigma_{\mathcal{R}_{\lambda,\rm mod}} ^2 \right )^{1/2}$ (the error on the observed, $\sigma_{\mathcal{R}_\lambda}$, and model FRS, $\sigma_{\mathcal{R}_{\lambda,\rm mod}}$, is due to the error of the best-fitting PSD normalisation, and were computed using eq. (\ref{eq:frs_def}) and (\ref{eq:frs_transfer}), respectively). We chose as our best-fitting, the model which resulted in the minimum $\chi^2$ ($\chi^2_{\rm min}$).

The best-fitting values for the corona height and accretion rate were: $h=2.5 \text{ }R_g$ and $\dot{m}_{\rm Edd}=0.25$\%, in the case of a maximally rotating BH ($\chi^2$ was systematically smaller in the case of the models with spin $\alpha=0.99$). The dashed red line in the top panel of Fig.\,\ref{fig:frs} indicates the best-fitting model. The best-fitting residuals are plotted in the bottom panel of the same figure. Clearly the model does not fit the data well ($\chi^2_{min}=81$ for 7 dof). The residual plot indicates that the best-fitting model overestimates the FRS values at short wavelengths and underestimates it at longer wavelengths.

Then, we considered the possibility that the UV/optical emission is further absorbed by the interstellar matter in the host galaxy of the active nucleus. We added an extra interstellar reddening component as a free parameter in our modelling. The fit is improved, but it is still not statistically acceptable ($\chi_{min}^2=34$ for 6 dof). The largest amplitude residuals appear in the $u$ and $r$ bands. This could be due to the the fact that up to 20\% of the observed flux in these bands may be due to Balmer continuum and H$\alpha$ emission, respectively \citep{2016ApJ...821...56F}. Both of these Broad Line Region (BLR) emission components are variable, and they could contribute to the observed FRS at these wavebands (estimated at a frequency which corresponds to a time-scale of 10 days). We therefore repeated the fit, ignoring the $u$ and $r-$band points. 

The fit is significantly improved ($\chi_{min}^2=7$, for 4 dof). The best-fitting model is shown by the blue solid line in the top panel of Fig. \ref{fig:frs}, and the best-fitting residuals are plotted in the middle panel of the same figure. The best fit parameters are: $h=5 \text{ }R_g$ ($2.5-20 \text{ }R_g$), $\dot{m}_{\rm Edd}=0.5$\% ($0.25-10$\%)\footnote{The confidence interval of the accretion rate is equal to the whole range of considered values, indicating that this parameter is mostly unconstrained by the fit. } and $E(B-V)=0.086 \text{ }(0.022-0.17)$; with the numbers in parenthesis indicating the $3\sigma$ confidence interval for the model parameters.


\section{Discussion}
\label{sec:discus}

We used the multi-wavelength data of NGC 5548 that were collected during the STORM campaign a few years ago, and we performed a power-spectrum analysis over a very broad energy range, from X-rays to UV and optical. Thanks to the unprecedented quality of the STORM light curves, we were able to estimate the PSD accurately over a broad range of time-scales, from $\sim 2.5$ up to $\sim 50$ days. To the best of our knowledge, this is the first time (for an AGN) that a power-spectrum analysis has been performed using simultaneous light curves, from X-rays to the optical band, covering a broad range of time-scales (from days to weeks). Our results can be summarised as follows.

We found that the X-ray PSD, is well fitted by a simple power law, with a slope of around $ -1.6$. This is almost identical to the results of \cite{2003ApJ...593...96M} who used {\it RXTE} and {\it XMM-Newton} data and estimated the X-ray PSD from 1/0.1 to $\sim 1/1000$ day$^{-1}$. They found that a simple power-law model with a slope of around $ -1.65$ can fit the PSD well. A broken power-law model could also fit well the PSD, with a break frequency of $\nu_b = 6.31 \cdot 10^{-7} (\mathrm{Hz} \simeq 0.055 \text{ }\mathrm{day}^{-1}$), although the improvement to the fit in the case of the broken power law model was not statistically significant. This frequency is close to the lowest frequency we probe in this work and, as a result, we cannot detect a break. We fitted the HX PSD with a broken power-law model, but the quality of the fit did not improve significantly, and the best-fitting parameters were entirely unconstrained. 

All the UV/optical PSDs have a power-law shape with a common slope consistent to $\sim -2$ (with the far-UV PSDs being slightly steeper). This apparent similarity is unlikely to be a coincidence and, most probably, suggests that the same mechanism drives the variability in all wavebands and over the frequency range considered here. On the other hand, the UV/optical PSD normalisation increases as a power law from the longer to the shorter wavelengths. In fact, it is this PSD normalisation dependence on energy that determines the variability amplitude decrease with increasing wavelength. In a way, such a decrease in the variability amplitude is expected due to the fact that the disc emission at longer wavelengths originates from regions with significantly larger areas. 

The X-ray PSD normalisation is much larger than the normalisation of the UV/optical PSDs, and its slope is flatter. The larger amplitude of the X-ray variations is consistent with the assumption that X-rays are emitted from a much smaller region in comparison to the disc. In addition, the smaller amplitude of the UV/optical variations  at short time-scales is consistent with X-ray reprocessing of the disc. In this case, the high-frequency variations of the X-ray source are expected to be smoothed out as the corresponding emission is reprocessed by the much larger region of the accretion disc.

\subsection{Indication of UV non-sationarity}

The P1 and P2 PSD normalisation of the  {\it Swift}/UVOT W1 and W2 light curves are significantly different (Fig.\,\ref{fig:psd_1}). The variability amplitude  in these two filters is much stronger during period P1. The PSD normalisation difference is significant at more than the 3-$\sigma$ level (Table \ref{tab:best-fit}).This is a strong indication for non-stationarity in the UV emission of NGC 5548.

The fact that the PSD best-fitting slope does not change suggests that the variability mechanism (responsible for the observed variability at the probed time-scales) remains the same during the two periods. A change in the variability amplitude can be explained within the X-ray reprocessing model, if there is a change in the source geometry while the physical processes remain the same. For example, according to KPD19, the amplitude of the thermally reprocessed disc flux increases with increasing corona height (see the plots in the right panel of their Fig.\,1). If the corona height was larger in period P1, the flux of the variable UV component would increase. This would imply a higher variability amplitude with respect to the total mean flux assuming a constant disc accretion rate. In fact, if we look in the middle and bottom panels in Fig.\,\ref{fig:lc_1}, the strong W2 and W1 flux increase  during period P1 (after $t\sim 6450$ day) results into a peak with a flux larger than the flux at any time during the period of the STORM observations. 

Variations in the height of the corona above the BH was also proposed by \cite{2019MNRAS.482.2088A} in order to explain the lack of stationary in IRAS 13224-3809, while \cite{2019Natur.565..198K} concluded that a reduction in the corona's size above the BH may have taken place in the case of the black-hole transient MAXI J1820+070. In this case, the change in the corona size would imply a change in the average height of the corona from the disc. \cite{2019MNRAS.482.2088A} detected significant, albeit of small amplitude, changes in the X-ray PSD of IRAS 13224-3809 at high frequencies. In our case the X-ray PSDs in periods P1 and P2 are consistent, within the errors. Perhaps, variations of the thermally reprocessed disc emission amplitude are much stronger than the changes in the X-ray PSD that the corona height variations may cause.

\subsection{Physical interpretation of FRS results}
\label{sec:frs_disc}

We studied the energy dependence of the UV/optical variability amplitude by means of Fourier Resolved Spectroscopy. FRS analysis can be a powerful tool to constrain variability models and the inner geometry of AGN. Recently, \cite{2020ApJ...891..178S} proposed to explain the UV/optical AGN variability with magnetic fluctuations which are produced close to the BH and are propagated outwards at Alfv\'en velocity. Their model is not consistent with our results. For instance, the model PSD in the optical band (in the case when the magnetic fluctuations have a power spectrum of $P_{mf} \propto \nu ^{-1}$) is steeper than what we observe. The predicted variability amplitude also decreases with wavelength slower than observed. The model predicts a variability amplitude ratio at 3000 \AA \ over  5100 \AA \  less than 2, while we observed a difference larger than  $\sim 3$ (Figure \,\ref{fig:frs}).

We found that the FRS at frequency $\nu = 0.1 \text{ } \mathrm{day}^{-1}$ has a power-law shape, with a slope of $\sim -2$.  According to \citep{2007ApJ...661...38P} this could be the case if the variable component has a power-law shape with the same slope and varies in normalisation only. However, it is not straightforward to attribute such a power-law wavelength dependence to a specific physical emission mechanism in the UV/optical band. 
As we already discussed in \S \ref{sec:frs_model} this result could not be due to a single BB component, with a variable temperature, in the inner disc. Another possibility would be coherent, global variations of the disc emission spectrum. According to the ``standard" \cite{1973A&A....24..337S} model, the disc should emit like a multi-temperature BB, and its emission should have a power-law like shape with a slope of around $-2.3$ in the UV/optical band. This is a bit steeper than the best-fitting FRS slope (it is just consistent within the $3\sigma$ upper limit). Nevertheless, global, instantaneous accretion rate variations are not possible, but perhaps propagating accretion rate variations could explain the observed FRS. But in this case, it would not be possible to explain the observed time lags. 

The observed FRS can be well reproduced in the case of X-ray reprocessing by the disc. We used the response functions of \elias and the X-ray PSD, and we found that the observed FRS can be explained if the X-ray corona is located at about $\sim 5 R_g$ above the central BH, and if we assume an intrinsic reddening of  $E(B-V)\simeq 0.08$ mag (consistent with the \cite{1998ApJ...499..719K} result of $E(B-V)=0.07^{+0.09}_{-0.06}$ mag, due to NLR absorption of the central source). Just like \elias, our results suggest that the observed UV/optical variability in NGC 5548 is the result of X-ray reprocessing in the disc. However, our results are not consistent with those reported by \elias. In particular, our best-fitting corona height is smaller than the KPD19 estimate. Although the two values agree at the 3$\sigma$ level, the difference is quite large. 
The time-lags (which KPD19 studied) depend on the width of the response function, while the PSD depends on the response amplitude. As long as the model tries to increase the corona height to explain the large time-lags (this was the case with the KPD19 results), the flux of the variable component, as well as the PSD amplitude, should also increase (contrary to the observed FRS). 

One possibility could be that the inner disc does not contribute to the observed UV/optical emission. For example, \cite{2015A&A...575A..22M} studied the broad band SED of NGC 5548 and concluded that the inner disc may be covered by a warm, optically-thick corona. Similar situation may hold in other objects as well. For instance, \cite{2018A&A...609A..42P, 2019A&A...623A..11P} used  broad-band spectral modelling and found strong indications for the presence of a warm, optically-thick corona, which may be extended, in Ark 120. This may have relevance to the observed time lags, which were reported by \cite{2020MNRAS.494.1165L} to be larger than expected from standard disc theory (while remaining consistent with the $\tau \propto \lambda^{4/3}$ relation, similar to NGC 5548). If such a warm corona, with a large optical depth exists, then all the UV/optical photons emitted by the underlying disc will be scattered to soft X-rays. In this case, the disc response functions will start at later times, and the overall flux of the X-ray reverberating component will decrease. Preliminary work to model the effects of a truncated disc indicated a better agreement of a large corona height with the observed FRS. A more thorough study is needed to evaluate the exact effect of such a layer in both the FRS and the time lags, before a conclusion can be reached.


\section{Conclusions}

In the present work, we studied the power spectra of NGC 5548 emission in different wavebands, from optical to X-rays, using data from the 2014 STORM monitoring campaign. To the best of our knowledge, this is the first time that such a broadband power spectral analysis is conducted for an AGN. We summarise below the main findings of our work:

\begin{enumerate}
  \item All the PSD were well described by a simple power law model. The X-ray PSD is consistent with previous results, while the UV/optical PSDs have the same slope, larger than the slope of the X-ray PSD.
  \item We found strong evidence of non-stationarity in the UV. The amplitude of the W2 and W1 PSDs differs significantly with time. This could be an indication that the disc/corona geometry varies on time scales of months/years in this object.
  \item The energy dependence of the observed variability amplitudes was investigated by studying the Fourier-Resolved Spectrum in the UV/optical band. The FRS has a power-law like shape, but it is not consistent with the hypothesis of a BB variable emission. On the other hand, the FRS is fully consistent with the hypothesis of disc thermal reverberation, due to X-ray illumination by a small X-ray corona.
\end{enumerate}

We plan to perform a multi-wavelength power-spectrum analysis using the observed light curves from the recent {\it Swift} monitoring campaigns of a few AGN \citep[e.g., NGC 4151, NGC 4593, Mrk 509,][]{2018MNRAS.480.2881M, 2019ApJ...870..123E}. We also plan to further develop the model fitting method, based on eq.\,\ref{eq:conv_fourier}. In principle, we can fit the full PSDs, at all wavelengths, using the model disc response functions but this is a complicated and time consuming approach. We plan to improve and update our modelling methods along this line in a future publication.


\section*{Data Availability}

The data underlying this article will be shared on reasonable request to the corresponding author.



\bibliographystyle{mnras}
\bibliography{paper_5548} 




\onecolumn{

\appendix

\section{Estimation of an FRS for a Blackbody}

\label{ap:bb_frs}

Let us consider a body with constant surface area, which emits as a BB with a variable temperature,  $T=T(t)$. Its spectrum (in units of $W\cdot sr^{-1}\cdot m^{-3}$) is  given by the Planck function:

\begin{eqnarray}
      B_\lambda(T) = \frac{2hc^2}{\lambda^5} \frac{1}{e^{hc/\lambda kT}-1},
\end{eqnarray}

\noindent where $c$ is the speed of light, $h$ and $k$ are the Planck and Boltzmann constants, respectively. If the temperature is variable, then the flux, at each wavelength, is also variable. Let us suppose we want to predict the FRS, $\mathcal{R}_{BB}(\lambda,\nu)$, in this case. According to eq. \ref{eq:frs_def} (Section \ref{sec:frs}):

\begin{eqnarray}
\label{eq:frs_def2}
  \centering
      \mathcal{R}_{BB}(\lambda,\nu) = <B_\lambda(T)>\sqrt{P(\lambda,\nu) d\nu},
\end{eqnarray}

\noindent where $<B_\lambda(T)>$ is the mean, at wavelength $\lambda$, and the PSD, $P(\lambda,\nu)$, is normalized to the mean flux squared. To compute the model PSD, we need to estimate the autocovariance function of the variable process at lag $\tau$, 

\begin{eqnarray}
  \centering
      ACF(\lambda,\tau) = E\left \{ \left [B_\lambda(T) - <B_\lambda(T)>\right ] \left [B_\lambda(T)(t+\tau) - <B_\lambda(T)>\right ]\right \},
\end{eqnarray}

\noindent where $E$ is the expectation operator. For a real-valued and stationary process, PSD and ACF are related by the equation.

\begin{eqnarray}
\label{eq:psd_acf}
  \centering
      P(\lambda,\nu) = \int_{-\infty}^{+\infty} cos(2\pi \nu\tau)ACF(\lambda,\tau) d\tau.
\end{eqnarray}
 
\noindent Let us write the variable temperature as: $T(t) = <T> + T_{V}(t)$, where $<T>$ is the mean temperature, and let us assume that the the amplitude of temperature variations is small, i.e. $\frac{|T_{V}(t)|}{<T>} \ll 1$. In this case, the following equations are valid:

\begin{eqnarray}
  \centering
      \frac{hc}{\lambda kT}= \left ( \frac{\lambda k <T>}{hc} + \frac{\lambda k T_V}{hc} \right )^{-1} \simeq \frac{hc}{\lambda k <T>} - \frac{hc}{\lambda k} \frac{T_V}{<T>^{2}},
\end{eqnarray}

\begin{eqnarray}
  \centering
      e^{{hc}/{\lambda kT}} \simeq e^{{hc}/{\lambda k<T>}} \cdot e^{{-hcT_V}/{\lambda k<T>^2}} \simeq e^{{hc}/{\lambda k<T>}} \left (1 - \frac{hc}{\lambda k} \frac{T_V}{<T>^{2}} \right ),
\end{eqnarray}

\begin{eqnarray}
  \centering
      \frac{1}{e^{{hc}/{\lambda kT}}-1} \simeq \frac{1}{e^{{hc}/{\lambda k<T>}} - \frac{hc}{\lambda k} \frac{T_V}{<T>^{2}} \cdot e^{{hc}/{\lambda k<T>}} - 1} \simeq \frac{1}{e^{{hc}/{\lambda k<T>}}-1} + \frac{\frac{hc}{\lambda k <T>} \frac{T_V}{<T>} \cdot e^{{hc}/{\lambda k<T>}}}{\left (e^{{hc}/{\lambda k<T>}} -1 \right )^2}.
\end{eqnarray}

\noindent Using the above three equations we can write:

\begin{eqnarray}
  \centering
      <B_\lambda(T)> = \frac{2hc^2}{\lambda^5} \frac{1}{e^{{hc}/{\lambda k<T>}}-1},
\end{eqnarray}

\noindent and,

\begin{eqnarray}
  \centering
      B_\lambda(T) - <B_\lambda(T)> = \frac{2hc^2}{\lambda^5} \frac{hc}{\lambda k<T>^2} e^{{hc}/{\lambda k<T>}} \frac{T_V}{\left (e^{{hc}/{\lambda k<T>}}-1 \right)^2}.
\end{eqnarray}

\noindent Hence, the ACF of a BB with a variable temperature is given by:

\begin{eqnarray}
  \centering
      ACF_{BB}(\lambda,\tau) = \left [\frac{2h^2c^3}{\lambda^6k<T>^2} \frac{e^{hc/\lambda k<T>}}{\left (e^{hc/\lambda k<T>}-1 \right )^2} \right ]^2 ACF_{T_{V}}({\tau}),
\end{eqnarray}

\noindent where $ACF_{T_{V}}(\tau)$ is the ACF of the process that is responsible for the temperature variations. The Fourier transform of $ACF_{BB}(\lambda, \tau)$ (which gives the power spectrum, $P(\lambda,\nu)$) is equal to the Fourier transform of $ACF_{T_{V}}(\tau)$ (times a constant), which is equal to the PSD of the mechanism responsible for the temperature variations, $P_{T_{V}}(\nu)$. Using then eq.\,\ref{eq:frs_def2}, we can finally get the FRS of a BB which is variable in temperature: 

\begin{eqnarray}
\label{eq:frs_bb}
  \centering
      \mathcal{R}_{BB}(\lambda,\nu) = \frac{2h^2c^3}{\lambda^6k<T>^2} \frac{e^{hc/\lambda k<T>}}{\left (e^{hc/\lambda k<T>}-1 \right )^2} \sqrt{P_{T_{V}}(\nu) d\nu}.
\end{eqnarray}

\noindent A plot  of $\mathcal{R}_{BB}(\lambda,\nu)$ as a function of the wavelength (at any frequency), for an arbitrary value of $P_{T_{V}},$ is shown in Fig. \ref{fig:frs_bb}. Different values of $P_{T_{V}}$ would modify the normalisation of this function. The plot is shown for a BB with a mean temperature of 10$^5$ K. A different  average temperature will merely shift the peak of the curve. At long wavelengths, $\mathcal{R}_{BB}(\lambda,\nu)$ follows the same power law as a BB emission, that is a power law with a slope of -4. It is evident from the plot, that the model FRS does not follow a power law with slope -2 in any part of the spectrum. We note that, following a similar approach, one may show that the FRS of a BB emission which varies only in normalisation will feature the same shape as the BB emission law.

\begin{figure}
  \centering
  \includegraphics*[width=0.5\columnwidth, clip]{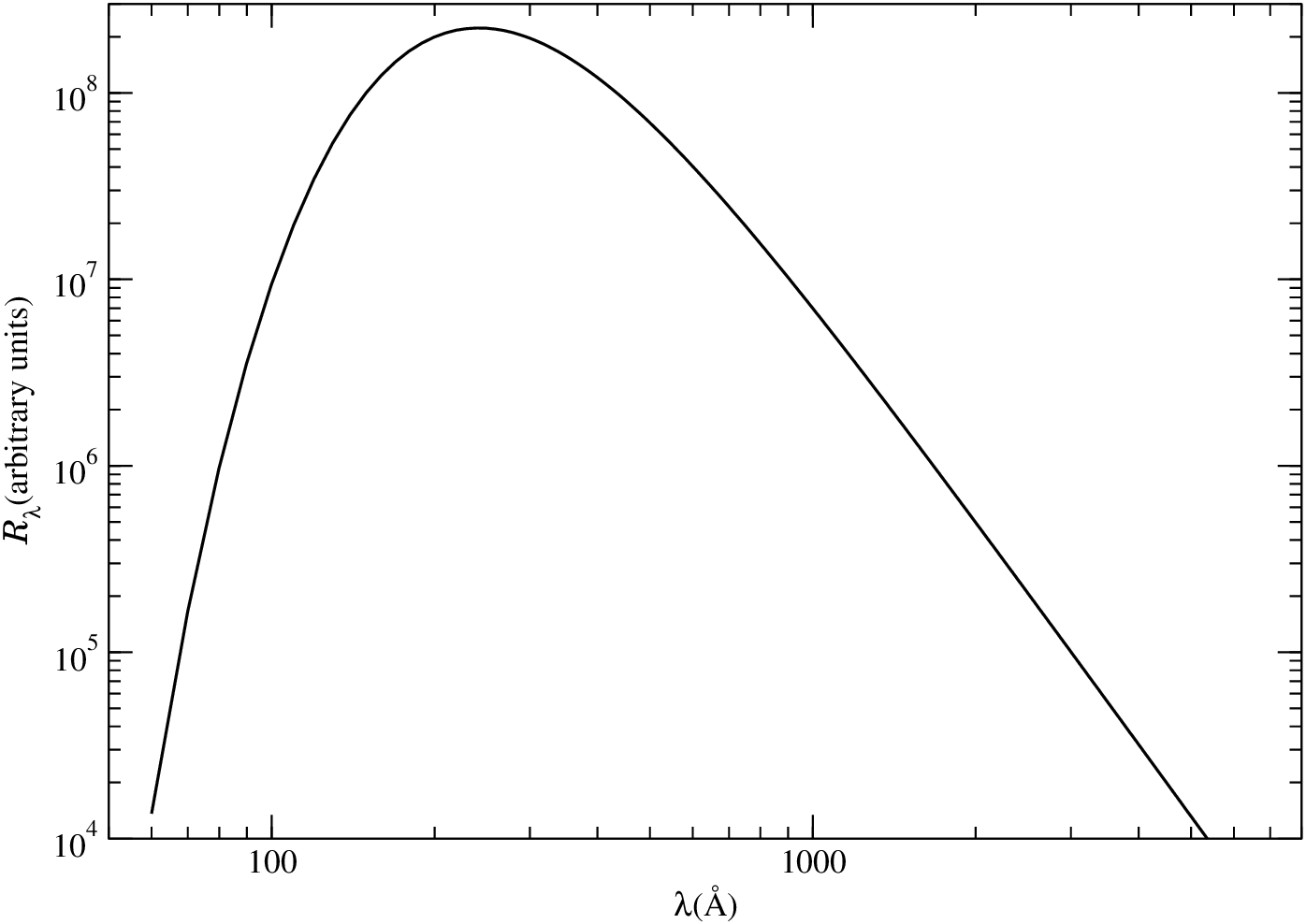}
  \caption[10]{The predicted FRS for a BB emission with variable temperature and an average temperature of 100000 K.}
  \label{fig:frs_bb}
\end{figure}

}

\bsp	
\label{lastpage}
\end{document}